\documentclass[aps,pra,twocolumn,showpacs,amsmath,amssymb]{revtex4}

\usepackage{color}
\usepackage{graphicx}
\usepackage{mathtools}

\begin{document}

\title{Particle-Hole Pair Coherence in Mott Insulator Quench Dynamics}

\author{K. W. Mahmud,$^1$ L. Jiang,$^1$ P. R. Johnson,$^2$ and E. Tiesinga$^1$}

\affiliation{$^1$Joint Quantum Institute, National Institute of
Standards and Technology and University of Maryland, 100 Bureau
Drive, Mail Stop 8423, Gaithersburg, Maryland 20899, USA\\
$^2$Department of Physics, American University, Washington, DC
20016, USA}

\begin{abstract}
We predict the existence of novel collapse and revival
oscillations that are a distinctive signature of the short-range
off-diagonal coherence associated with particle-hole pairs in Mott
insulator states. Starting with an atomic Mott state in a
one-dimensional optical lattice, suddenly raising the lattice
depth freezes the particle-hole pairs in place and induces phase
oscillations. The peak of the quasi-momentum distribution,
revealed through time of flight interference, oscillates between a
maximum occupation at zero quasi-momentum (the $\Gamma$ point) and
the edge of the Brillouin zone. We show that the population
enhancements at the edge of the Brillouin zone is due to coherent
particle-hole pairs, and we find similar effects for fermions and
Bose-Fermi mixtures in a lattice. Our results open a new avenue
for probing strongly correlated many-body states with short-range
phase coherence that goes beyond the familiar collapse and
revivals previously observed in the long-range coherent superfluid
regime.
\end{abstract}


\maketitle

\begin{figure}[b]
\vspace{-0.2cm}
\begin{center}
  \includegraphics[width=0.4\textwidth,angle=0]{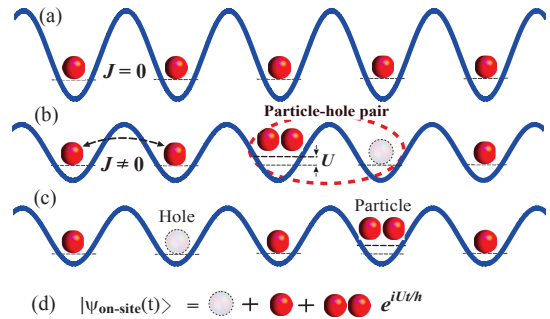}
\end{center}
\vspace{-0.7cm} \caption{\label{fig:sketch} (color online). A
schematic of the Mott insulator state and quench. (a) A perfect
Mott state with zero tunneling ($J=0$). For small tunneling $J
\neq 0$, there exist nearest-neighbor and next-nearest-neighbor
particle-hole pairs, as seen in (b) and (c), respectively. A
sudden raise of the lattice depth freezes the atoms in place, and
induces phase evolution of the on-site wavefunctions as
schematically depicted in (d).}
\end{figure}

Ultracold atoms in optical lattices are a versatile tool for
creating strongly correlated quantum many-body
states~\cite{duchon13,bloch08}. Prominent examples include atomic
Mott insulator (MI) states of bosons and
fermions~\cite{greiner02a,jordens08,schneider08}. Starting from a
superfluid (SF) with long-range phase coherence, adiabatically
increasing the ratio of atom-atom interaction $U$ to tunneling
strength $J$ by increasing the optical lattice depth gives rise to
a MI with short-range coherence~\cite{greiner02a}. An infinitely
deep lattice with $J=0$ gives a \textit{perfect} Mott insulator
state as depicted in Fig.~\ref{fig:sketch}(a), denoted in Fock
space notation as $|11..11\rangle$ for unit occupancy. For finite
tunneling but still in the MI regime, the many-body wavefunction
includes correlated (paired) double and zero occupied sites on top
of the perfect MI [see Fig.~\ref{fig:sketch}(b), (c)]. These
particle-hole pairs, also known as doublons and holons and denoted
as $|201..1\rangle, |210..1\rangle$, etc., behave as
quasi-particle excitations. Particle-hole pairs play an important
role in Mott insulator physics~\cite{gerbier05,bakr09,bakr10}, and
have recently been observed by {\it in-situ} imaging in a
one-dimensional (1D) chain of bosonic atoms~\cite{endres11}.

Ultracold atoms in optical lattices are also a versatile tool for
studying many-body quantum
dynamics~\cite{kennett13,polkovnikov11}. A common technique
involves quenching a system by suddenly changing its
parameters~\cite{kennett13,kollath07}. One approach, starting from
larger $U/J$ and quenching to a smaller $U/J$, has been used to
study thermalization~\cite{altman02,rigol08}, the Kibble-Zurek
mechanism~\cite{zurek05,demarco11}, and light-cone spreading of
correlations~\cite{lightcone12,barmettler13,natu12}. The opposite
approach, starting in a SF regime with small $U/J$ and quenching
to large $U/J$ gives the collapse-and-revival (CR) oscillations of
matter-wave phase coherence observed in
Refs.~\cite{greiner02b,porto07,will10}. Matter-wave CR has found
applications in the study of higher-body
interactions~\cite{will10,mahmud13a}, fermion
impurities~\cite{will11}, and coherent-state
squeezing~\cite{will10,tiesinga11}. To date, CR has been viewed as
a characteristic behavior of quenched systems with long-range
phase coherence such as superfluids.

\begin{figure*}[ht]
\vspace{-0.2cm}
\begin{center}
  \includegraphics[width=0.75\textwidth,angle=0]{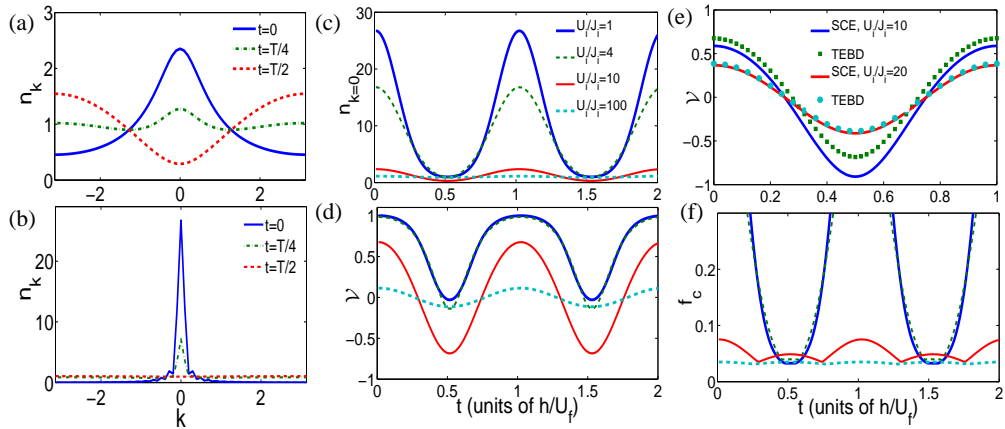}
\end{center}
\vspace{-0.8cm} \caption{\label{fig:nkdynamics} (color online).
Numerical simulations of Bosonic Mott-insulator coherence
dynamics. (a) The quasi-momentum population $n_k$ as a function of
$k$ for hold times $t=0, T/4$ and $T/2$, for a MI with occupation
${\bar n}=1$ and $U_i/J_i=10$. During the collapse of the $k=0$
occupation, a revival occurs at the Brillouin zone edge at
$k=\pi$. In contrast, (b) shows a SF quench for ${\bar n}=1$ and
$U_i/J_i=1$, where a collapse of the $k=0$ population is not
accompanied by a revival at $k=\pi$. (c) $n_{k=0}(t)$ as a
function of hold time for four values of $U_i/J_i$, either in the
SF or MI regimes, and ${\bar n}=1$. (d) Visibility ${\cal V}(t)$
dynamics for the same values of $U_i/J_i$ as in (c) have a
sinusoidal form and reach negative values (indicating
$n_{k=\pi}>n_{k=0}$) for MI states but not for SF. (e) A
comparison between the strong-coupling expansion (SCE) theory and
numerical i-TEBD results shows good agreement deep in the MI
regime (e.g., $U_i/J_i=20$) but not closer to the phase boundary
(e.g., $U_i/J_i=10$). (f) The condensate fraction $f_c(t)$
dynamics, for the four $U_i/J_i$ given in (c), shows novel kinks
in the deep MI regime. }
\end{figure*}

In this Letter, we show that collapse-and-revival also occurs for
quenched MI states, for both lattice bosons, fermions, and mixtures.
Surprisingly, detectable CR oscillations should occur despite the
short-range coherence of the Mott state.
For a 1D optical lattice, the quasi-momentum
distribution oscillates between maximum occupations (revivals) at
quasi-momentum $\hbar k=0$ and at the edge of the Brillouin zone.
(Here $\hbar=h/(2 \pi)$ and $h$ is Planck's constant.) In fact,
the normalized difference in the two populations, which is
conventionally defined as the ``visibility" of the condensate,
becomes negative, in sharp contrast to the behavior of quenched
superfluids. We also find that the visibility of the quenched MI
state is a sinusoidal function of time, in contrast to the
exponential functional form seen for quenched superfluids. This
difference provides another approach for distinguishing between
systems with short- and long-range coherence.

For a quenched MI, the condensate fraction oscillations also show novel
kinks associated with oscillations between symmetric and antisymmetric
natural orbitals of the single-particle density matrix (SPDM). The
distinctive behavior of MI CR oscillations are due to the presence of
correlated particle-hole pairs, frozen in place when the lattice
depth is suddenly increased, leading to phase dynamics as schematically
illustrated in Fig.~\ref{fig:sketch}(d).

Unexpectedly, we find that fermions in a lattice at half-filling,
and Bose-Fermi mixtures when the total occupation of bosons plus
fermions is an integer, exhibit similar dynamics due to correlated
particle-hole pairs. Finally, we show that this physics is robust
under realistic dephasing mechanisms, and should be within
experimental reach. Our results reveal universal coherence
dynamics for quenched, strongly-correlated lattice systems with
short-range phase coherence.

\emph{Lattice bosons.}$-$ Ultracold bosons in an optical lattice
can be described by the Bose-Hubbard Hamiltonian
\begin{eqnarray}
H^{{\rm B}}_{i}=-J_{i}\sum_{\langle jj' \rangle}\left(
b_{j}^{\dagger}b_{j'}+ {\rm h.c.} \right) +
\dfrac{U_{i}}{2}\sum_{j}n_{j}\left(n_{j}-1\right),
\label{eqn:boseHubb}
\end{eqnarray}
where $j, j'$ are indices to lattice sites, only nearest-neighbor
tunneling is assumed, and $n_j= b_{j}^{\dagger}b_{j}$. The
subscript $i$ denotes parameter values before the quench. The
total number of particles is $N=\sum_{j} \langle
b_{j}^{\dagger}b_{j}\rangle$, and ${\bar n}=N/L$ is the mean
occupation per site where $L$ is the number of sites. For
non-integer occupation per site, the ground state is superfluid.
At integer occupation, the system exhibits a quantum phase
transition, going from a SF to MI state above a critical value of
$\eta_c=U/J$. At unit occupation, $\eta_c \approx 3.37$ in
1D~\cite{scalettar91,zakrzewski08}.

The system is quenched by suddenly increasing the depth of the
optical lattice such that tunneling is suppressed ($J_f
\rightarrow 0$)~\cite{greiner02b}. The Hamiltonian governing the
post-quench dynamics is
$H_f=(U_f/2)\sum_{j}n_{j}(n_{j}-1)$,
where $U_{f}$ is the final interaction strength. After a
hold-time $t$, the lattice is turned off and the atoms are allowed
to freely expand. Imaging the atoms after the time-of-flight
expansion, the atomic spatial density corresponds to the
quasi-momentum distribution $n_{k}(t)
=(1/L)\sum_{j,j'}e^{ik(j-j')}\rho_{jj'}(t)$ at the moment of
release, where $\rho_{jj'}(t)=\langle b_{j}^{\dagger}b_{j'}\rangle$
is the SPDM and $k$ is the lattice wavevector. In a 1D lattice
with a period of unit length, the edge of the Brillouin zone is at
wavevector $k=\pi$. In addition, we examine the visibility
${\cal V}(t)=[n_{k=0}(t)-n_{k=\pi}(t)]/[n_{k=0}(t)+n_{k=\pi}(t)]$~\cite{gerbier05},
and the condensate fraction $f_c(t)=\lambda_c(t)/N$, where
$\lambda_c(t)$ is the largest eigenvalue of the SPDM \cite{penrose56,hofstetter11}.

In previous CR studies the initial state is a superfluid.  Here, our
initial state is a one-dimensional MI  with integer ${\bar n}$
and $U_i/J_i > \eta_c$. Due to the presence of particle-hole pairs,
this state, after the quench, is not an eigenstate of $H_f$, and
undergoes nonequilibrium evolution until the moment of release from the
lattice. Much of the MI quench dynamics can be understood from an analysis
of correlations in the initial state. It is convenient to define the
off-diagonal coherence $\zeta_d(t)=\langle b_{j}^{\dagger}b_{j+d}\rangle$,
which is independent of $j$ as the SPDM $\rho_{jj'}$ only depends on
$|j-j'|$ for a homogeneous system with periodic boundary conditions.
It follows that $n_{k=0}(t)={\bar n} +2\sum_{d>0} (1-d/L) \zeta_d(t)$ and,
similarly, $n_{k=\pi}(t)={\bar n} +2\sum_{d>0} (-1)^d (1-d/L) \zeta_d(t)$.

We are able to obtain explicit analytic approximations for the evolution
of the off-diagonal coherence and quasimomentum populations for bosons
in a 1D lattice using the strong-coupling expansion \cite{freericks96}
up to second-order in the tunneling Hamiltonian.  The full derivation
is given in the Supplementary material. Briefly, to zeroth-order
the ground-state wavefunction is the Mott state
$|0\rangle =|{\bar n},{\bar n},...,{\bar n}\rangle$ with ${\bar n}$ atoms
in each of the $L$ sites.  Corrections are due to states containing
a ``hole'', ${\bar n}-1$ atoms in one of the sites, and a ``particle'',
${\bar n}+1$ atoms in another. We can express these particle-hole
pair states as $|e_d\rangle \propto \sum^{L}_{j=1}(b^{\dagger}_j
b_{j+d}+b^{\dagger}_{j+d} b_{j})|0\rangle$, where $d>0$ is the lattice
distance between particle and hole. The improved ground state becomes
$|g \rangle=|0\rangle + \sum_{d>0} \epsilon_d|e_d\rangle$ with $\epsilon_d\propto
(J_i/U_i)^d\ll 1$, from which it follows that  initially $\zeta_d(t=0)\propto (J_i/U_i)^d$.
This corresponds to an exponential decay of the off-diagonal coherence with
distance $d$~(see also Ref.~\cite{cazalilla11}).

Deep in the MI regime the time evolution of the $k=0$ and $k=\pi$
populations are therefore, to good approximation, $n_{k=0}(t)
\approx 1+2(1-1/L) \zeta_1(t)$ and $n_{k=\pi}(t) \approx
1-2(1-1/L) \zeta_1(t)$, respectively, for a MI with ${\bar n}=1$.
Consequently, when $\zeta_1(t)$ is positive (negative) the
population $n_{k=0}(t)$ is larger (smaller) than $n_{k=\pi}(t)$.
In fact, it is shown in the Supplementary material that
$\zeta_1(t) \propto (J_i/U_i)\cos(U_f t/\hbar)$  with period
$T=h/U_f$. Hence, $n_{k=\pi}(t)$ is larger than $n_{k=0}(t)$ for
$T/2<t<3T/2$. This implies that ${\cal V}(t)\propto
n_{k=0}(t)-n_{k=\pi}(t)<0$ for these hold times.

In contrast, for a SF with $U_i/J_i<\eta_c$ for ${\bar n}=1$ and
with any $U_i/J_i$ for non-integer $\bar n$, the mean-field
ground-state wavefunction is site separable, i.e., of the form
$\prod_j |S_j\rangle$, with $|S_j\rangle=\sum_{n} c^{(j)}_n
|n\rangle$. Consequently, the SPMD is also separable with
$\zeta_d=\langle b^\dagger_j\rangle\langle b_{j+d}\rangle$. For a
homogeneous lattice $\zeta_d=|\langle b\rangle|^{2}$ is
independent of $d$ and there is no decay of correlation with
particle-hole distance $d$. As all $\zeta_d$ are positive we find
$n_{k=0}(t) \geq n_{k=\pi}(t)$ for all $t$. This implies that
${\cal V}(t)>0$. This conclusion also holds for a fully correlated
model when superfluid SPDM is not constant but decays with $d$
algebraically in 1D and goes to a constant value in higher
dimensions. Due to the slower than exponential decay, all the
terms in the expression of $n_{k}(t)$ contribute, and $n_{k=0}(t)
\geq n_{k=\pi}(t)$.

In the Supplementary material we also derive the quasi-momentum and
visibility in the MI regime and the thermodynamic limit. They are given by
\begin{eqnarray}
\lefteqn{
n_{k}(t) \approx  {\bar n}+4{\bar n}({\bar n}+1)\frac{J_i}{U_i}\cos\left(\frac{U_f t}{\hbar}\right) \cos(k)}
\label{eqn:nk} \\
&& + 2{\bar n}({\bar n}+1)(2{\bar
n}+1)\frac{J^{2}_i}{U^{2}_i}\left[1+2\cos\left(\frac{U_f
t}{\hbar}\right)\right] \cos(2k), \nonumber
\end{eqnarray}
and
\begin{eqnarray}
{\cal V}(t) &\approx &4({\bar n}+1)\frac{J_i}{U_i}\cos(U_f t/\hbar).
\label{eqn:nu1}
\end{eqnarray}
In contrast,
in the superfluid regime with a homogeneous pre-quench state that is separable,
$n_{k=0}(t)=(L-1) v(t)+{\bar n}$ and ${\cal
V}(t)=v(t)/(v(t)+2[{\bar n}-v(t)]/L)$, where $v(t)=|\langle b(t)\rangle|^2$.
For a coherent state $|\beta\rangle$ at each lattice site
\begin{equation}
v(t)=|\langle b(t)\rangle|^{2}={\bar n} e^{2 {\bar n}\left[
{\cos}(U_f t/\hbar)-1\right]}, \label{eqn:SFvisib}
\end{equation}
where ${\bar n}=|\beta|^{2}$ and
$b_{j}(t)=e^{iH_{f}t/\hbar}b_{j}e^{-iH_{f}t/\hbar}=e^{-iU_f
n_{j}t/\hbar}b_{j}$.
For both the MI and SF quench
the time evolution is periodic with period $T=h/U_f$.
The oscillations, however, are sinusoidal for the MI quench and ``exponential'' for the SF quench.

We next perform numerical i-TEBD simulations \cite{vidal07} for a
MI with ${\bar n}=1$ to extend our analysis to all values of
$U_i/J_i$ and validate the SCE analysis in the MI regime. We
calculate the SPDM for $L=32$ and use periodic boundary
conditions. Figure~\ref{fig:nkdynamics}(a) shows the
quasi-momentum $n_k(t)$ for a Mott state with $U_i/J_i=10$ at hold
times $t=0$, $T/4$, and $T/2$. At $t=0$ the momentum distribution
is peaked around $k=0$, while at $t=T/2$ it is peaked around
$k=\pi$, corresponding to a ``revival'' at the edge of the
Brillouin zone and to a new kind of nonequilibrium state. This
$n_{k=\pi}(t)$ revival is a distinctive signature of the presence
of nearest-neighbor particle-hole pairs in the MI state.

To highlight the difference between MI and superfluid quenches,
Fig.~\ref{fig:nkdynamics}(b) shows the $n_k(t)$ dynamics for a
superfluid with $U_i/J_i=1$. Here, at $t=0$ the quasimomentum
distribution is again peaked around $k=0$, but is much narrower
than the MI case due to the long-range coherence of the SF. During
the collapse at $t=T/2$, the atoms are equally spread over all
quasi-momenta with no enhancement or revival of the $k=\pi$
population. Note that $n_{k=\pi}(t)$ revivals have also been
predicted for supersolid quenches in a system with (long-range)
nearest-neighbor interactions~\cite{fischer12}. Those revivals,
however, are associated with long-range coherences and not
particle-hole pairs.

Figure~\ref{fig:nkdynamics}(c) and (d) show the time evolution of
$n_{k=0}(t)$ and ${\cal V}(t)$ for different values of $U_i/J_i$
and ${\bar n}=1$, one in the SF regime and three in the MI regime. For
$U_i/J_i=1$, the time trace exhibits superfluid CR oscillations consistent
with Eq.~\ref{eqn:SFvisib} and the visibility is positive. Based on our
numerical results, we conjecture that for $U_i/J_i>\eta_c$ there will always exist
time intervals where the visibility is negative. We show an example for
$U_i/J_i=4$ in Fig.~\ref{fig:nkdynamics}(d).
For larger $U_i/J_i$ the oscillations in $n_{k=0}(t)$ and ${\cal
V}(t)$ become more sinusoidal, consistent with Eqs.~\ref{eqn:nk} and
\ref{eqn:nu1}. Examples are given for $U_i/J_i=10$ and $100$.

\begin{figure}
\vspace{-0.2cm}
\begin{center}
  \includegraphics[width=0.45\textwidth,angle=0]{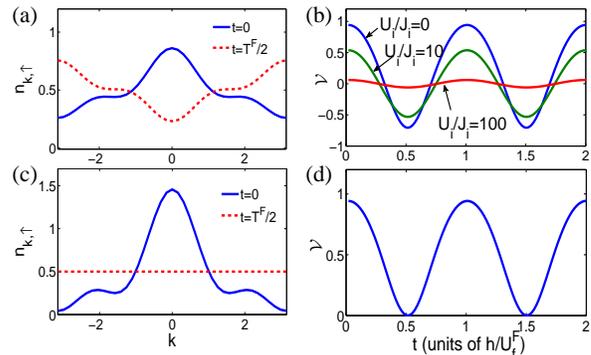}
\end{center}
\vspace{-0.9cm} \caption{\label{fig:fermion} (color online).
Lattice fermions. (a) $n_{k,\uparrow}$ as a function of $k$ at
times $t=0$ and $T^{\rm F}/2$ for a half-filled state. (b) The
visibility dynamics of the spin-up state shows sinusoidal
oscillations for several $U_i/J_i$. Panels (c) and (d) show
$n_{k,\uparrow}$ versus $k$ and spin-up visibility versus time, respectively, of a
non-half-filled metallic state, with superfluid like CR.}
\end{figure}

In Fig.~\ref{fig:nkdynamics}(e) we compare analytic predictions for
${\cal V}(t)$ in Eq.~\ref{eqn:nu1} with our numerical simulations. The
analytic approximation gives excellent agreement for $U_i/J_i \gtrsim 20$.
The regime of validity for the analytic theory can be extended
to smaller $U_i/J_i$ by performing higher-order calculations and keeping
high occupancy terms, e.g., terms such as $|..030..\rangle$ and $|..2020..\rangle$
for unit occupancy ($\bar n=1$).

Figure~\ref{fig:nkdynamics}(f) shows the condensate fraction
$f_c(t)=\lambda_c(t)/N$ as a function of time.  For $U_i/J_i$ in the MI
regime $f_c(t)$ has kinks at times $t=T/4$ and $3T/4$, which occur when
the eigenfunction or natural orbital of the SPDM associated with the
largest eigenvalue changes between a symmetric state $(1,1,..1,1)$ and
an anti-symmetric state $(1,-1,..,1,-1)$.  This behavior is consistent
with our SCE calculations, which give
\begin{eqnarray}
\lambda_{c}(t) &=&{\bar n}+4{\bar n}({\bar n}+1)\frac{J_i}{U_i}|\cos\left(U_f t/\hbar\right)| \label{eqn:fc}\\
&&+2{\bar n}({\bar n}+1)(2{\bar
n}+1)\frac{J^{2}_i}{U^{2}_i}[1+2\cos\left(U_f t/\hbar\right)]\,,
\nonumber
\end{eqnarray}
and where the second term leads to the kinks.  Within the SCE
approximation the kinks occur when the nearest neighbor coherence
$\zeta_{d=1}$ goes to zero.  For $U_i/J_i=1$ in the SF regime the
fraction performs smooth oscillations reaching $1/N$. (The full
range is not shown in the graph.) The numerical result is
consistent with $f_c(t)=v(t)/{\bar n}+({\bar n}-v(t))/N$, based on
the mean-field approximation \cite{hofstetter11}.

\begin{figure}[t]
\vspace{-0.0cm}
\begin{center}
  \includegraphics[width=0.45\textwidth,angle=0]{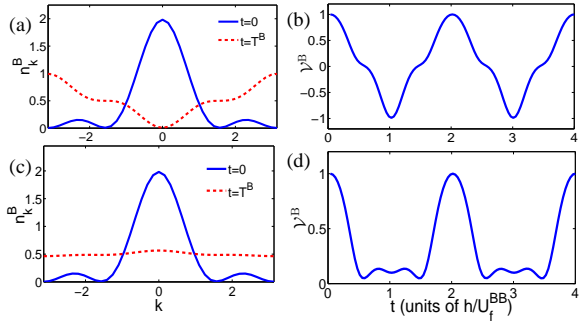}
\end{center}
\vspace{-0.8cm} \caption{\label{fig:bosefermi} (color online).
Lattice Bose-Fermi mixtures. (a) $n^{\rm B}_{k}$ as a function of
$k$ at times $t=0$ and $T^{\rm B}$, with $T^{\rm B}=h/U^{\rm
BB}_{f}$ and $U^{\rm BF}_f/U^{\rm BB}_f=0.5$, and a combined unit
occupation of bosons and fermions. A maximum in the $k=\pi$
occupation occurs when the $k=0$ occupation is near zero
(collapses). (b) The bosonic visibility shows MI-like sinusoidal
features. In contrast, a state with non-integer combined
occupation gives SF-like CR dynamics in $n^{\rm B}_k$ (c) and
bosonic visibility (d).}
\end{figure}

\begin{figure} [t]
\vspace{-0.0cm}
\begin{center}
  \includegraphics[width=0.45\textwidth,angle=0]{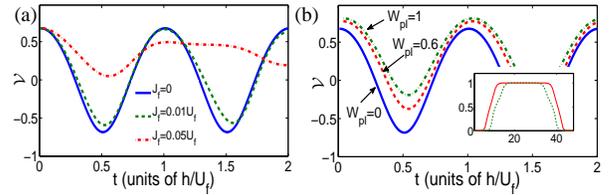}
\end{center}
\vspace{-0.8cm} \caption{\label{fig:detection} (color online).
Effects of residual tunneling and initial harmonic confinement on
bosonic MI dynamics. (a) Visibility dynamics for three values of
$J_f$. Other parameters as in Fig.~\ref{fig:nkdynamics}(a). (b)
Visibility dynamics for an initially trapped system for
$W_{pl}=0,0.6$, and $1$, where $W_{pl}$ is the ratio of the width
of SF regions to the Mott plateau. Inset shows the initial density
profile as a function of lattice site.}
\end{figure}

\emph{Lattice fermions.}$-$ Fermions in an optical lattice are
described by the Fermi-Hubbard Hamiltonian
\begin{eqnarray}
H^{{\rm F}}_{i}=-J^{{\rm F}}_{i}\sum_{\langle jj' \rangle
\sigma}\left( f_{j\sigma}^{\dagger}f_{j'\sigma}+ {\rm h.c.}
\right)+U^{{\rm F}}_{i}\sum_{j}n_{j\uparrow}n_{j\downarrow},
\label{eqn:fermiHubb}
\end{eqnarray}
where $\sigma=\uparrow,\downarrow$ denotes two spin states and
$n_{j\sigma}=f_{j\sigma}^{\dagger}f_{j\sigma}$. Its phase diagram
contains metallic, Mott insulator and anti-ferromagnatic
phases~\cite{lee06}. After the quench the Hamiltonian becomes
$H^{{\rm F}}_{f}=U^{{\rm
F}}_{f}\sum_{j}n_{j\uparrow}n_{j\downarrow}$.
Figure~\ref{fig:fermion} illustrates the quench dynamics, computed
using a four site lattice and exact diagonalization. We first
consider the half-filled case with $\langle
n_{j\uparrow}\rangle=\langle n_{j\downarrow}\rangle=1/2$.
Figure~\ref{fig:fermion}(a) shows the momentum distribution
$n_{k,\uparrow}$ at times $t=0$ and $T^{\rm F}/2$, where $T^{\rm
F}=h/U^{{\rm F}}_f$. At $t=0$ the momentum distribution is peaked
around $k=0$, while at $t=T^{\rm F}/2$ the peak occurs at the edge
of the Brillouin zone, an effect similar to the bosonic MI case.
The $k=\pi$ revivals are again due to correlated doublons and
holons in the initial state. Visibility oscillations of the
spin-up state, shown in Fig.~\ref{fig:fermion}(b), reach negative
values for any $U_i/J_i$ and become more sinusoidal for larger
values of $U_i/J_i$. In contrast, Figs.~\ref{fig:fermion}(c) and
(d) show that a metallic state with $\langle
n_{j\uparrow}\rangle=1/2$  and $\langle
n_{j\downarrow}\rangle=1/4$ exhibits the familiar superfluid-type
CR, where all quasi-momenta are uniformly occupied during collapse
at $t=T^{\rm F}/2$.

\emph{Lattice Bose-Fermi mixtures.}$-$ For Bose-Fermi mixtures in
a lattice, the post-quench Hamiltonian is
\begin{eqnarray}
H^{{\rm BF}}_{f}=\frac{U^{{\rm BB}}_{f}}{2}\sum_{j}n^{{\rm B}}_{j}
\left( n^{{\rm B}}_{j}-1 \right)+ U^{{\rm BF}}_{f}\sum_{j}n^{{\rm
B}}_{j}n^{{\rm F}}_{j}, \label{eqn:bosefermi}
\end{eqnarray}
where $U^{{\rm BB}}_f$ and $U^{{\rm BF}}_f$ are Bose-Bose and
Bose-Fermi interaction strengths, respectively, and assuming fermions in a single spin state. For the pre-quench
Hamiltonian, the usual hopping terms are also
present~\cite{albus03}. Bose-Fermi mixtures give rise to a
plethora of phases and phenomena~\cite{albus03,anders12},
depending on the relative interaction strengths and fermion
filling. CR of a Bose-Fermi mixture has been observed in
Ref.~\cite{will11}. When the total occupation per site of fermions
plus bosons is one, the ground state phase diagram has a
MI state~\cite{altman12}. The quench dynamics, computed with a
four site lattice, is shown in Figs.~\ref{fig:bosefermi}(a) and
(b). Here again we see $k=\pi$ revivals of the bosonic momentum
distribution $n_k$ and the sinusoidal behavior of the bosonic
visibility, with modifications due to the Bose-Fermi interaction.
In contrast, for non-unit occupation, the dynamics gives the
SF-like CR shown in Figs.~\ref{fig:bosefermi}(c) and (d).

\emph{Experimental prospects and conclusions.}$-$ Our analysis in
Fig.~2 shows that the predicted visibility of the $k=\pi$ revival
should be of the same order of magnitude as the $k=0$ peak
observed for an equilibrium MI~\cite{gerbier05}, and hence a
detection of the dynamics should be possible. These effects are
also robust under several dephasing mechanisms. First,
Fig.~\ref{fig:detection}(a) shows that for small post-quench tunneling
$J_f$ the characteristic negative ${\cal V}$ survives, but is destroyed
for larger $J_f/U_f$.
Second, we find that effective multi-body
interactions~\cite{will10,johnson09,johnson12} do not significantly
influence the dynamics deep in the ${\bar n=1}$ MI regime (e.g.,
$U_i/J_i> 10$). Third, by adding $V_{T,i} \sum_{j} (j-L/2)^2 n_j$
in Eq.~\ref{eqn:boseHubb}, harmonically trapped systems give a MI
core with SF regions at the two edges. Assuming $V_{T,f}=0$ after
the quench, Fig.~\ref{fig:detection}(b) shows that the SF regions
modify ${\cal V}(t)$, but do not completely wash out the
distinctive MI signature for a sufficiently large Mott plateau.
Dephasing due to post-quench harmonic trapping $V_{T,f} \neq 0$
for SF revivals has been quantified in
Refs.~\cite{hofstetter11,mahmud13b}.

In conclusion, we have shown that bosons and fermions in 1D
periodic potentials with short-range coherence will manifest a new
type of collapse-and-revival oscillation due to particle-hole pair
correlations. We expect qualitatively similar effects in 2D and 3D.
Our results open up the possibility that the underlying correlations of
a wide class of strongly correlated matter can now be revealed through
coherence dynamics.

\emph{Acknowledgments.}$-$ We acknowledge support from the US Army
Research Office under Contract No. 60661PH.

\clearpage

\section{Supplementary Material}
\setcounter{equation}{0}

We derive analytic results for quench dynamics of a
one-dimensional Mott insulator using the strong-coupling expansion
\cite{freericks96}.  Bosons in an optical lattice are described by
the Bose-Hubbard Hamiltonian of Eq.~\ref{eqn:boseHubb}.  We assume
a homogeneous system with $L=2M+1$ sites, periodic boundary
conditions, and atom number $N={\bar n}L$, where $M$ and ${\bar
n}$ are positive integers.

We prepare the Mott-insulator ground state assuming a non-zero
tunneling energy $J_i$ that is small compared to the atom-atom
interaction strength $U_i$. Following \cite{freericks96} we
perform perturbation theory in the tunneling operator or kinetic
energy.  To zeroth-order the ground state wavefunction is the Mott
state $|0\rangle =|{\bar n},{\bar n},...,{\bar n}\rangle$ with
${\bar n}$ atoms in each of the $L$ sites.  The  correction to
this wavefunction is due to states containing a single hole with
$n\!-\!1$ atoms in one of the sites and a single ``particle'' with
$n+1$ atoms in another.  We classify these (normalized)
particle-hole pair states as
\begin{eqnarray*}
|e_{d}\rangle  &=&\frac{1}{\sqrt{2L}} ( \dots + |...,{\bar
n}\underbracket[1pt]{-1,...,{\bar n}+}_{d\ \rm sites}1,...\rangle+
         ... \\
&&\quad \quad\quad \dots + |...,{\bar
n}\underbracket[1pt]{+1,...,{\bar n}-}_{d\ \rm sites}1,...\rangle+
...)
       \,,
\end{eqnarray*}
where $d=1,\dots,M$ is the number of sites that separates the hole
and particle.  The dots in a ket indicate sites with ${\bar n}$
atoms and the sum is over all states with the same particle and
hole separation. The hole is either to the left or right of the
particle. (Periodic boundary conditions imply that the largest
separation is $M$.) The Bose-Hubbard Hamiltonian within the space
spanned by $\{|0\rangle ,|e_{1}\rangle ,...|e_{M}\rangle \}$ is
given by
\begin{eqnarray*}
&&H = \left[
\begin{array}{cc}
E_{0} & A^T \\
A & K
\end{array}
\right]\,,
\end{eqnarray*}
where $E_{0}=U_iL {\bar n}({\bar n}-1)/2$ is the diagonal matrix
element for state $|0\rangle$, $A^T=(-\sqrt{2L{\bar n}({\bar
n}+1)}J_i,0,0,0,...)$ with $M$ elements describes the coupling
between $|e_{d}\rangle$ and $|0\rangle$, and
\begin{equation*}
K=\left[
\begin{array}{cccc}
E_{0}+U_i  & -(2{\bar n}+1)J_i &     0    &  ... \\
-(2{\bar n}+1)J_i &  E_{0}+U_i & -(2{\bar n}+1)J_i &  ... \\
    0    & -(2{\bar n}+1)J_i &  E_{0}+U_i &  ... \\
   ...   &    ...   &    ...   & ...
\end{array}
\right]
\end{equation*}
is a $M\times M$ tridiagonal (diagonal-constant) matrix that
describes the couplings among the $|e_{d}\rangle$.

All diagonal elements of $K$ are the same, and we must first
diagonalize $K$ in order to perform perturbation theory. Its
eigenenergies and functions are
\begin{equation*}
{\cal E}_{k}=E_{0}+U_i-2(2{\bar n}+1)J_i\cos \left(\frac{\pi
k}{M+1}\right)
\end{equation*}
and
\begin{equation*}
|{\tilde e}_{k}\rangle = \sqrt{\frac{2}{M+1}} \sum_{j=1}^M \sin
\left(\frac{\pi jk}{M+1} \right) |e_{d}\rangle \,.
\end{equation*}

Performing perturbation theory with respect to the non-degenerate
states $|{\tilde e}_{k}\rangle$, the correction to the ground state
wavefunction is
\begin{eqnarray*}
|g\rangle &=& \left(1-L{\bar n}({\bar
n}+1)\frac{J_i^{2}}{U_i^{2}}\right)|0\rangle
+\sqrt{2L{\bar n}({\bar n}+1)}\frac{J_i}{U_i}|e_{1}\rangle \\
&&+(2{\bar n}+1)\sqrt{2L{\bar n}({\bar
n}+1)}\frac{J_i^{2}}{U_i^{2}}|e_{2}\rangle + O[(J_i/U_i)^3]
\end{eqnarray*}
in terms of the original particle-hole basis.

We now quench the system by suddenly setting the tunneling to zero
and change $U_i\to U_f$. The subsequent time evolution is then
given by
\begin{eqnarray}
|g(t)\rangle &= & (1-L{\bar n}({\bar n}+1)\frac{J_i^{2}}{U_i^{2}})|0\rangle  \label{eq:timedep}\\
       && \quad
+\sqrt{2L{\bar n}({\bar n}+1)}\frac{J_i}{U_i}|e_{1}\rangle e^{-iU_ft/\hbar} \nonumber\\
       && \quad
    +(2{\bar n}+1)\sqrt{2L{\bar n}({\bar n}+1)}\frac{J_i^{2}}{U_i^{2}}|e_{2}\rangle e^{-iU_ft/\hbar}
           \nonumber
\end{eqnarray}
as the states $|0\rangle$ and $|e_d\rangle$ are eigenstates of the
quenched Hamiltonian.  In fact, the $|e_d\rangle$ are degenerate
with an energy $U_f$ relative to the state $|0\rangle$.

Our experimental observables are the quasi-momentum distribution
at the $\Gamma$ point and the edge of the Brillouin zone, defined
as $n_{k=0}(t) =\langle g(t)|b_{k=0}^{\dagger}b_{k=0}|g(t)\rangle
=\sum_{jj'}\rho _{jj'}(t)/L$, and $n_{k=\pi }(t)
=\sum_{jj'}(-1)^{j-j'}\rho _{jj'}(t)/L$, respectively. Here $\rho
_{jj'}(t)=\langle g(t)|b_{j}^{\dagger}b_{j'}|g(t)\rangle$ is the
$L\times L$ single-particle density matrix. Finally, the
visibility is defined by
\begin{eqnarray*}
      {\cal V}(t)=\frac{n_{k=0}(t)-n_{k=\pi }(t)}{n_{k=0}(t)+n_{k=\pi }(t)} \,.
\end{eqnarray*}

The single-particle density matrix can be evaluated using
Eq.~\ref{eq:timedep} and leads to
\begin{eqnarray*}
\rho _{jj'}(t) &=& \left(1-2L{\bar n}({\bar
n}+1)\frac{J_i^{2}}{U_i^{2}}\right)
                                       \langle 0|b_{j}^{\dagger}b_{j'}|0\rangle \\
            & & +\sqrt{2L{\bar n}({\bar n}+1)}\frac{J_i}{U_i}
                   \left(\langle 0|b_{j}^{\dagger}b_{j'}|e_{1}\rangle e^{-iU_ft/\hbar}\right.  \\
            & & \phantom{+\sqrt{2L{\bar n}({\bar n}+1)}\frac{J_i}{U_i} (}
                \left. \quad\quad
                        + \langle e_{1}|b_{j}^{\dagger}b_{j'}|0\rangle e^{iU_ft/\hbar}\right) \\
            & & +(2{\bar n}+1)\sqrt{2L{\bar n}({\bar n}+1)}\frac{J_i^{2}}{U_i^{2}} \\
            & &  \quad \times
                   \left(\langle 0|b_{j}^{\dagger}b_{j'}|e_{2}\rangle e^{-iU_ft/\hbar}
                         +\langle e_{2}|b_{j}^{\dagger}b_{j'}|0\rangle e^{iU_ft/\hbar}\right) \\
            & &
                 +2L{\bar n}({\bar n}+1)\frac{J_i^{2}}{U_i^{2}}\langle e_{1}|b_{j}^{\dagger}b_{j'}|e_{1}\rangle
                  \,,
\end{eqnarray*}
to second order in $J_i/U_i$. Further analysis shows that
$\rho_{jj'}(t)=\zeta_{|j-j'|}(t)$ is a positive-definite pentadiagonal
matrix with diagonal-constant coefficients given by $\zeta_0(t)=n$,
$\zeta_1(t)=2{\bar n}({\bar n}+1)(J_i/U_i)\cos(U_ft/\hbar)$,
$\zeta_2(t)={\bar n}({\bar n}+1)(2{\bar
n}+1)(J_i^{2}/U_i^{2})(1+2\cos (U_ft/\hbar))$, and zero otherwise.
In the thermodynamic limit, the quasi-momentum dynamics is given
by
\begin{eqnarray*}
&n_{k}(t)& = {\bar n}+4{\bar n}({\bar n}+1)\frac{J_i}{U_i}\cos\left(\frac{U_f t}{\hbar}\right) \cos(k) +\nonumber\\
&& 2{\bar n}({\bar n}+1)(2{\bar
n}+1)\frac{J^{2}_i}{U^{2}_i}\left[1+2\cos\left(\frac{U_f
t}{\hbar}\right)\right] \cos(2k),
\end{eqnarray*}
and the visibility is
\begin{eqnarray*}
{\cal V}(t) &=&\frac{4{\bar n}({\bar n}+1)\frac{J_i}{U_i}\cos
(U_ft/\hbar)}{{\bar n}+2{\bar n}({\bar n}+1)(2{\bar
n}+1)\frac{J_i^{2}}{U_i^{2}}[1+2\cos U_ft/\hbar]}
\\
&\approx &4({\bar n}+1)\frac{J_i}{U_i}\cos (U_ft/\hbar) \,.
\end{eqnarray*}

We find that it is also useful to define $\lambda_{c}(t)$, the
largest eigenvalue of the single-particle density matrix at time
$t$.  As $\rho_{jj'}(t)$ is a real symmetric matrix, we
can use perturbation theory to find its largest
eigenvalue.  We define $\rho(t)=\rho^{(0)}(t)+\delta\rho(t)$ with
tri-diagonal matrix $\rho^{(0)}(t)$ given by the diagonal and sub-
and super-diagonal of $\rho(t)$.  The matrix
 $\rho^{(0)}(t)$ can be diagonalized analytically and its largest eigenvalue is
\begin{eqnarray*}
\lambda^{(0)}_{c}(t) &=& {\bar n}+4{\bar n}({\bar
n}+1)\frac{J_i}{U_i}|\cos (U_ft/\hbar)|\cos \left(\frac{\pi
}{L+1}\right)\,,
\end{eqnarray*}
with corresponding eigenvector $\vec v$ and elements
\begin{eqnarray*}
v_j &=&\sqrt{\frac{2}{L+1}}
                      \sin \left(\frac{\pi j}{L+1}\right)
\end{eqnarray*}
when $U_ft/\hbar\in (-\pi/2+2\pi m ,\pi/2+2\pi m)$, and
\begin{eqnarray*}
v_j &=&\sqrt{\frac{2}{L+1}}(-1)^{j}\sin \left(\frac{\pi
j}{L+1}\right)\,,
\end{eqnarray*}
when $U_ft/\hbar\in (\pi/2+2\pi m,3\pi/2+2\pi m)$. Here $m$ is any
integer and index $j=1,\dots,L$.

Finally, the matrix $\delta\rho(t)$ corrects the largest eigenvalue
and we have
\begin{eqnarray*}
\lambda_{c}(t) &=&  \lambda^{(0)}_{c}(t)+ {\vec v}^T\cdot \delta\rho(t) \cdot {\vec v}\\
      &=&     {\bar n}+4{\bar n}({\bar n}+1)\frac{J_i}{U_i}|\cos (U_ft/\hbar)| \\
      & &   \quad
              +\, 2{\bar n}({\bar n}+1)(2{\bar n}+1)\frac{J_i^{2}}{U_i^{2}}[1+2\cos
              (U_ft/\hbar)] \,,
\end{eqnarray*}
to first order in $\delta\rho(t)$.

\end{document}